\documentclass[a4paper, 10pt]{article} 
\usepackage{graphicx, amssymb} 
\usepackage[sumlimits]{amsmath}
\usepackage{amsthm}
\usepackage{array}

\usepackage{fullpage}

\newtheorem{thm}{Theorem}
\newtheorem{lem}[thm]{Lemma}

\theoremstyle{remark}

\newcommand{\refeqn}[1]{(\ref{eqn:#1})}

\newcommand{\reflem}[1]{Lemma~\ref{lem:#1}}

\newcommand{\cD}{{\cal D}}

\newcommand{\cM}{{\cal M}}

\newcommand{\adv} {\mathop{\mathrm{Adv}^\pm}}

\newcommand{\s}[1]{\left(#1\right)}

\newcommand{\pfstart}{\begin{proof}} 
\newcommand{\pfend}{\end{proof}} 

\title{Adversary Lower Bound for Element Distinctness}
\author{Aleksandrs Belovs\thanks{Faculty of Computing, University of Latvia, stiboh@gmail.com.}}
\date{}

\begin{document}
\maketitle

\begin{abstract}
In this note we construct an explicit optimal (negative-weight) adversary matrix for the element distinctness problem, given that the size of the alphabet is sufficiently large.
\end{abstract}

\section{Introduction}
Two main techniques for proving lower bounds on quantum query complexity are the polynomial method~\cite{polynomial} developed by Beals {\em et al.} in 1998, and the adversary method~\cite{adversary} developed by Ambainis in 2000. Both techniques are incomparable. There are functions with adversary bound strictly larger than polynomial degree~\cite{polynomialVsQCC}, as well as functions with the reverse relation.

Two main examples of the reverse relation are exhibited by the collision and the element distinctness functions. The input to both functions is a string of length $n$ of symbols in an alphabet of size $q$, i.e., $x = (x_i)\in [q]^n$. We use notation $[q]$ to denote the set $\{1,\dots,q\}$. The element distinctness function evaluates to 0 if all symbols in the input string are pairwise distinct, and to 1 otherwise. The collision function is defined similarly with the additional promise that in the positive case the input can be divided into pairwise disjoint pairs of equal elements with different values in different pairs, i.e., for each $i\in [n]$ there exists unique $j\in [n]\setminus\{i\}$ such that $x_i=x_j$. Clearly, this problem is non-trivial only if $n$ is even.

The quantum query complexity of the collision and the element distinctness problems are $O(n^{1/3})$ and $O(n^{2/3})$, respectively, with the algorithms given by Brassard {\em et al.}~\cite{collision} and Ambainis~\cite{distinctness}, respectively. The tight lower bounds were given by Aaronson and Shi~\cite{collisionLowerShi}, Kutin~\cite{collisionLowerKutin} and Ambainis~\cite{collisionLowerAmbainis} using the polynomial method. 

The adversary bound, however, fails for these functions. The so-called certificate complexity barrier~\cite{adversaryEquivalent, adversaryPower} implies that the best lower bound achievable by the adversary method is mere $\Omega(\sqrt{n})$. Similarly, the so-called property testing barrier~\cite{adversaryNegative} shows that an $\omega(1)$ lower bound for collision is out of reach for the adversary method.

In 2006, a stronger version of the adversary bound was developed by H\o yer {\em et al.}~\cite{adversaryNegative}. This is the so-called negative-weight adversary lower bound. Later it was proved to be optimal by Reichardt {\em et al.}~\cite{adversaryTight1, adversaryTight2}. Let us define it.

Assume $f:\cD\to \{0,1\}$ is a function with domain $\cD\subseteq [q]^n$. Let $\cM$ be the set of non-zero real matrices with rows indexed by the elements of $f^{-1}(1)$, and columns indexed by the elements of $f^{-1}(0)$. For $i\in[n]$, let $\Delta_i$ be the matrix from $\cM$ with entry $(x,y)$ equal to 1 if $x_i\ne y_i$, and to 0 otherwise. The {\em negative-weight adversary bound} is defined as
\begin{equation}
\label{eqn:adversary}
\adv(f) = \max_{\Gamma\in \cM} \frac{\|\Gamma\|}{\max_{i\in[n]} \|\Gamma\circ \Delta_i\|},
\end{equation}
where $\|\cdot\|$ is the spectral norm, and $\circ$ is the Hadamard (element-wise) product of matrices.

Let $Q(f)$ denote the query complexity of the best quantum algorithm evaluating $f$ with a bounded error. Then we have the following result:
\begin{thm}[\cite{adversaryNegative, adversaryTight1, adversaryTight2}]
\label{thm:adversary}
Let $f$ be as above.  Then, $Q(f)=\Theta(\adv(f))$.
\end{thm}

Although the negative-weight adversary lower bound is known to be tight, it has been almost never used to prove lower bounds for explicit functions. Vast majority of lower bounds by the adversary method used the old positive-weight version of this method. But since the only competing polynomial method is known to be non-tight, a better understanding of the negative-weight adversary method would be very beneficial. In the sequel, we consider the negative-weight adversary bound only, and will omit the adjective ``negative-weight''.

The aforementioned proof of the lower bound for element distinctness using the polynomial method is circuitous, and consists of three steps:
\begin{enumerate}\itemsep0pt
\item an $\Omega(n^{1/3})$ lower bound on the collision problem (with somewhat large range) using polynomial method~\cite{collisionLowerShi, collisionLowerKutin};
\item a reduction from the the $\Omega(n^{1/3})$ collision lower bound to an $\Omega(n^{2/3})$ element distinctness lower bound with large range;
\item a reduction of the size of the range~\cite{collisionLowerAmbainis}. Unfortunately, this reduction only works for lower bounds obtained using the polynomial method.
\end{enumerate}
In this note, we construct an explicit adversary matrix that combines the first two steps. In other words, it gives an $\Omega(n^{2/3})$ lower bound for element distinctness, provided that the size of the alphabet $q=\Omega(n^2)$. We hope it will lead to proving adversary lower bounds for quantum query complexity of other functions.

\section{Construction}
The aim of this section is to prove the following result:
\begin{thm}
If $f$ is the element distinctness function with $q=\Omega(n^2)$, then there exists $\Gamma\in\cM$ such that the expression in~\refeqn{adversary} is $\Omega(n^{2/3})$.
\end{thm}

We may assume the rows of $\Gamma$ are indexed with the inputs with a unique collision, i.e., for  each index $x\in [q]^n$ of a row in $\cM$, there exist $a,b\in [n]$ such that $a\ne b$, $x_a=x_b$, and $x_i\ne x_j$ for all $i\ne j$ unless $\{i,j\}=\{a,b\}$. We say $\{a,b\}$ is the position of the collision in $x$. The columns of $\Gamma$ are indexed by inputs with all elements distinct.

The idea of our construction is to embed $\Gamma$ into a slightly larger matrix $\Gamma'\in \cM'$, where $\cM'$ is $\cM$ with additional rows and columns to be defined later. Define $\Delta_i$ on $\cM'$ in the same way it is defined on $\cM$. Then $\Gamma\circ \Delta_i$ is a submatrix of $\Gamma'\circ \Delta_i$, hence, $\|\Gamma\circ \Delta_i\|\le \|\Gamma'\circ \Delta_i\|$. The thing to prove then is that $\|\Gamma'\|$ is large, and that $\|\Gamma\|$ is not much smaller than $\|\Gamma'\|$.

The matrix $\Gamma'$ consists of ${n\choose 2}$ matrices $G_{a,b}$ stacked one on another for all possible choices $\{a,b\}\subset[n]$. That is,
\[\Gamma' = \begin{pmatrix}G_{1,2}\\G_{1,3}\\\vdots\\G_{n-1,n}\end{pmatrix}. \]
In $G_{a,b}$, the columns are labeled by the elements $y\in [q]^n$, and the rows are labeled by the elements $x\in[q]^n$ with the additional requirement $x_a=x_b$ (thus, there are $q^{n-1}$ rows in each $G_{a,b}$).

We say a column with index $y$ is {\em illegal }if $y_i=y_j$ for some $i\ne j$. Similarly, we say a row in $G_{a,b}$ with index $x$ is illegal if $x_i=x_j$ for some $i\ne j$ such that $\{i,j\}\ne\{a,b\}$. Thus, after removing all illegal rows and columns, $G_{a,b}$ will represent the part of $\Gamma$ with the rows indexed by the inputs having collision in $\{a,b\}$.

Clearly, since there are elements of $[q]^n$ that are used as labels of both rows and columns in $\cM'$, it is easy to design a matrix $\Gamma'\in \cM'$ such that the value in~\refeqn{adversary} is arbitrarily large. But we design $\Gamma'$ in such way that it still is a good adversary matrix after the illegal rows and columns are removed. 

Let $J_q$ be the $q\times q$ all-ones matrix. Assume $e_0,\dots,e_{q-1}$ is an orthonormal eigenbasis of $J_q$ with $e_0=1/\sqrt{q}(1,\dots,1)$ being the eigenvalue $q$ eigenvector. Consider the vectors of the following form:
\begin{equation}
\label{eqn:v}
v = e_{v_1}\otimes e_{v_2}\otimes\cdots\otimes e_{v_n},
\end{equation}
where $v_i \in \{0,\dots,q-1\}$. These are eigenvectors of the Hamming Association Scheme on $[q]^n$. For a vector $v$ from~\refeqn{v}, the {\em weight} $|v|$ is defined as the number of non-zero entries in $(v_1,\dots,v_n)$. Let $E_k^{(n)}$, for $k=0,\dots,n$, be the orthogonal projector on the space spanned by the vectors from~\refeqn{v} having weight $k$. These are the projectors on the eigenspaces of the association scheme. Let us denote $E_i = E^{(1)}_i$ for $i=0,1$. These are $q\times q$ matrices. All entries of $E_0$ are equal to $1/q$, and the entries of $E_1$ are given by
\[
(E_1)_{xy} = \begin{cases}
1-1/q,& x=y;\\
-1/q,& x\ne y.
\end{cases}
\]

Elements $a$ and $b$ in $G_{a,b}$ should be treated differently from the rest of the elements. For them, we define two $q\times q^2$ operators $F_0$ and $F_1$. They are defined by taking the $\sqrt{q}$ multiples of $E^{(2)}_0$ and $E^{(2)}_1$ and considering only the rows consisting of two equal elements. The first operator is $F_0=e_0(e_0\otimes e_0)^*$. The second one is
\[
F_1 = \sum_{i\ne 0} e_i (e_0\otimes e_i + e_i\otimes e_0)^*.
\]
All entries of $F_0$ are equal to $q^{-3/2}$. The entry of $F_1$ on the intersection of a row with index $(x,x)$ and a column with index $(y_1,y_2)$ is
\[
\frac{1}{\sqrt{q}}\begin{cases}
2-2/q,& y_1=y_2=x;\\
1-2/q,& \mbox{$y_1=x$, or $y_2=x$, but not both;}\\
-2/q,& \mbox{$y_1\ne x$ and $y_2\ne x$.}
\end{cases}
\]

We are going to define $G_{1,2}$ as a linear combination of $F_\ell\otimes E_k^{(n-2)}$ with $\ell\in\{0,1\}$, $k\in\{0,\dots,n-2\}$. All other $G_{a,b}$s are obtained from $G_{1,2}$ by permuting the input variables so that $\{1,2\}$ gets moved to $\{a,b\}$.

We have to understand three things: how the coefficients of $F_\ell\otimes E_k^{(n-2)}$s contribute to the norm of $\Gamma'$, how $\Delta_i$ acts of these elements, and how they affect the norm $\Gamma'\circ\Delta_i$.

\paragraph{Norm of $\Gamma'$} Let $W(G)= (\Gamma')^*\Gamma'$, where $\Gamma'$ is obtained from $G_{1,2}=G$ as described above. It is a positive semidefinite matrix with rows and columns labeled by elements of $[q]^n$. We describe it in the basis from~\refeqn{v}. At first, we have
\begin{equation}
\label{eqn:Wfirst}
W(F_0\otimes E_k^{(n-2)}) = \frac{(n-k)(n-k-1)}{2}E_k^{(n)}.
\end{equation}
Here the coefficient is equal to the number of variants of picking a pair of components of $v$ from~\refeqn{v} equal to $e_0$, if $v$ has weight $k$. Also, we have
\[
v^*\; W(F_1\otimes E_k^{(n-2)})\; v' =
\begin{cases}
(k+1)(n-k-1),& \text{$v=v'$ and $|v|=k+1$\;;}\\
1, & \parbox{8cm}{$|v|=|v'|=k+1$ and $v$ can be obtained from $v'$ by exchanging one $e_0$ with one $e_j$ with $j\ne 0$\;;}\\
0,&\text{otherwise}.
\end{cases}
\]
For each $v$ of weight $k+1$, there are exactly $(k+1)(n-k-1)$ choices for $v'$ such that the second option holds. Thus,
\begin{equation}
\label{eqn:Wsecond}
\left\|W(F_1\otimes E_k^{(n-2)})\right\| = 2(k+1)(n-k-1),
\end{equation}
and $W(F_1\otimes E_k^{(n-2)})$ acts as the zero operator on the orthogonal complement of $E^{(n)}_{k+1}$. 

\paragraph{Action of $\Delta_1$} 
Firstly, our construction ensures that $\Gamma'$ is symmetric with respect to the permutation of variables. Thus, it suffices to estimate the norm of $\Gamma'\circ \Delta_1$. Next, it is known that $\|A\circ \Delta_1\| \le 2\|A\|$ for any $A\in \cM'$~\cite{adversaryTight2}. Using this, and in order to simplify calculations, we consider not the mapping $\Gamma'\mapsto \Gamma'\circ \Delta_1$, but some mapping $\Gamma'\stackrel{\Delta_1}{\longmapsto} \Gamma'_1$ such that $\Gamma'\circ\Delta_1 = \Gamma'_1\circ\Delta_1$. In other words, we are allowed to change arbitrarily entries $(x,y)$ with $x_1=y_1$.

Since $\Delta_1$ acts trivially on all input variables except the first one, it is enough to consider its action on $E_0$, $E_1$, $F_0$, and $F_1$. For the first two, we define:
\begin{equation}
\label{eqn:EDelta}
E_0\stackrel{\Delta_1}{\longmapsto}E_0 \qquad\mbox{and}\qquad E_1 \stackrel{\Delta_1}{\longmapsto}- E_0.
\end{equation}
For $F_i$, we embed it back into $E^{(2)}$, apply the transformation from~\refeqn{EDelta} and remove the unnecessary rows again. In the result:
\[
F_0\stackrel{\Delta_1}{\longmapsto} F_0\qquad\mbox{and}\qquad F_1\stackrel{\Delta_1}{\longmapsto} -F_0+\sum_{i\ne 0} e_i(e_0\otimes e_i)^*.
\]
Thus, if we take $F=F_0+F_1$, we have
\begin{equation}
\label{eqn:f}
F \stackrel{\Delta_1}{\longmapsto} \sum_{i\ne 0} e_i(e_0\otimes e_i)^*.
\end{equation}

We are going to construct $G_{1,2}$ as 
\begin{equation}
\label{eqn:G12}
G_{1,2}=\sum_k \alpha_k F\otimes E_k^{(n-2)}
\end{equation}
for some non-negative real $\alpha_k$s.

\paragraph{Norm of $\Gamma'_1$} The matrix $\Gamma'$ can be naturally divided into two submatrices: the first one formed by the matrices $G_{1,b}$, and the second one by $G_{a,b}$s with $a,b\ne 1$. Let $W_1$ and $W_2$ be defined similarly as $W$, but after applying $\Delta_1$ and for the first and the second submatrices, respectively. 

Let's start with the first submatrix. From~\refeqn{f}, we have
\begin{equation}
\label{eqn:W1}
W_1(F\otimes E_k^{(n-2)})  = (k+1) E_0\otimes E^{(n-1)}_{k+1}.
\end{equation}

For $W_2$, we may assume $\{a,b\}=\{n-1,n\}$, analyze the action of $\Delta_1$ on $E^{(n-2)}_k$s, and then apply element permutation on the last $n-1$ elements like in the construction of $\Gamma'$ from $G_{1,2}$.

From~\refeqn{EDelta}, we have:
\[
\sum_k \alpha_k E_k^{(n-2)} \stackrel{\Delta_1}{\longmapsto} E_0\otimes \sum_k (\alpha_k - \alpha_{k+1}) E_k^{(n-3)}.
\]
Using this, as well as~\refeqn{Wfirst} and~\refeqn{Wsecond}, modified to the last $n-1$ elements, we get
\begin{equation}
\label{eqn:W2}
\left\|W_2\s{F\otimes \sum_k \alpha_k E^{(n-2)}_k } \right \| = O(n^2)\max_k (\alpha_k - \alpha_{k+1})^2.
\end{equation}

\paragraph{Optimizing $\alpha_k$} Now we are able to choose $\alpha_k$ optimally. We assume that the norm of $\Gamma'$ is maximized for the right singular subspace $E_0^{(n)}$. Thus, from~\refeqn{Wfirst}, we have 
\begin{equation}
\label{eqn:GammaNorm}
\|\Gamma'\|^2 = \Omega(\alpha_0^2 n^2).
\end{equation}

From~\refeqn{W1} and~\refeqn{W2}, we have the following conditions on $\|\Gamma'_1\|=O(1)$:
\[
\alpha_k^2\le \frac{1}{k+1}\qquad\mbox{and}\qquad \alpha_{k}-\alpha_{k+1} \le \frac1n.
\]
Thus, the maximal possible value of $\alpha_0$ equals, up to a constant factor, to the minimal value of $1/\sqrt{r} +r/n$ over $r$. The minimum is $n^{-1/3}$ that is attained at $r = n^{2/3}$. Thus, by~\refeqn{GammaNorm}, the optimal value of $\|\Gamma'\|$ is $\Omega(n^{2/3})$.

\paragraph{Norm of $\Gamma$} 
In order to finish the proof, it remains to show that $\|\Gamma\|$ is not much smaller than $\|\Gamma'\|$, where $\Gamma$ is obtained from $\Gamma'$ by striking out the illegal rows and columns. This is the only place where we use the assumption $q=\Omega(n^2)$. We start with the following technical observation.
\begin{lem}
\label{lem:summa}
Let $\tilde E_1^{(k)}$ be the matrix $E_1^{\otimes k}$ with the rows and the columns having equal elements in their indices are removed. The sum of the elements of $\tilde E_1^{(k)}$ is non-negative.
\end{lem}

\pfstart
Because of the symmetry, it suffices to prove that the sum of the elements of the first row of $\tilde E_1^{(k)}$ is non-negative. We prove a slightly more general statement by induction on $k$. Assume $k\le \ell\le q$ and let $g(k,\ell,q)$ denote the sum of the elements of the first row of the matrix $(qI_\ell - J_\ell)^{\otimes k}$ after the rows and the columns with equal elements are removed. Here $I_\ell$ and $J_\ell$ are the $\ell\times\ell$ identity and all-ones matrices, respectively. We prove that $g(k,\ell,q)\ge 0$ for all choices of $k$, $\ell$ and $q$. Then the sum of the elements of the first row of $\tilde E_1^{(k)}$ equals $q^{-k} g(k,q,q)\ge 0$.

Assume the first row is labeled by the sequence $(1,2,\dots, k)$. It is easy to see that $g(0,\ell,q)=1$ and $g(1,\ell,q)=q-\ell$ both are non-negative. Now assume $k\ge 2$. Then the labels of the columns $(y_1,\dots,y_k)$ of the matrix can be divided into three parts:
\begin{itemize}\itemsep0pt
\item $y_1=1$. The contribution of the entries in these columns is $(q-1)g(k-1,\ell-1,q)$.
\item $y_1\in [\ell]\setminus[k]$. The contribution from these columns is $-(\ell-k)g(k-1,\ell-1,q)$.
\item $y_1\in [k]\setminus\{1\}$. Assume $y_1=2$ at first. Considering $\{y_3,\dots,y_k\}$ only, we get $g(k-2,\ell-1,q)$. There are $\ell-k+1$ choices of $y_2$, when all other elements of $y$ are fixed. There are $k-1$ choices of $y_1$. Thus, the total contribution of these columns is $(k-1)(\ell-k+1)g(k-2,\ell-1,q)$. 
\end{itemize}
Altogether:
\[
g(k,\ell,q) = (q-\ell+k-1)g(k-1,\ell-1,q) + (k-1)(\ell-k+1)g(k-2,\ell-1,q)\ge 0, 
\]
because all the multiples are non-negative by the inductive assumption.
\pfend

From our construction of $\Gamma'$, we know that the singular value of the right singular vector $e_0^{\otimes n}$ is $\Omega(n^{2/3})$, and the corresponding left singular vector is the normalized all-ones vector as well. Hence, the singular value equals the sum of the elements of $\Gamma'$ divided by $\sqrt{{n\choose 2}q^{2n-1}}$.

It suffices to prove that the sum of the entires of $G_{1,2}$ doesn't drop much after all illegal rows and columns are removed. Let $\widetilde{G}_{1,2}$ be $G_{1,2}$ with these rows and columns removed. Consider expression~\refeqn{G12}. The contribution from $\alpha_0 F_0\otimes E_0^{n-k}$ to the sum of all entries drop by at most a constant factor, because an $\Omega(1)$ fraction of the rows and columns of $G_{1,2}$ is legal (here we use that $q=\Omega(n^2)$).

All other terms in~\refeqn{G12} contribute 0 to the sum of the elements in $G_{1,2}$. But their contribution to the sum of the elements of $\widetilde{G}_{1,2}$ is non-negative. Indeed, consider the terms 
\begin{equation}
\label{eqn:terms}
 \alpha_k F_0\otimes E_{v_3}\otimes\cdots\otimes E_{v_n},\qquad \alpha_k \sum_{i\ne 0} e_i(e_i\otimes e_0)^*\otimes E_{v_3}\otimes\cdots\otimes E_{v_n}\quad\mbox{and}\quad \alpha_k \sum_{i\ne 0} e_i(e_0\otimes e_i)^* \otimes E_{v_3}\otimes\cdots\otimes E_{v_n}
\end{equation}
independently. Here, $v_3,\dots, v_n\in \{0,1\}$, and exactly $k$ of them are 1. 

In each of them, entry $(x,y)$ of $\widetilde{G}_{1,2}$ only depends on $x_i$ and $y_i$ where $v_i=1$, as well as $i=1$ and $i=2$ in the second and the third cases, respectively. The entry of $\widetilde{G}_{1,2}$ is proportional to the corresponding entry of $\tilde E_1^{(k)}$ or $\tilde E_1^{(k+1)}$, where the irrelevant elements of $x$ and $y$ are removed. Moreover, the number of entries of $\widetilde{G}_{1,2}$ corresponding to an entry of $\tilde E_1^{(k)}$ (or $\tilde E_1^{(k+1)}$) is the same for all entries. Hence, by \reflem{summa}, the contribution of the terms in~\refeqn{terms} is non-negative. Thus, the norm of $\Gamma$ still is $\Omega(n^{2/3})$. 

\section*{Acknowledgments}
I would like to thank Andris Ambainis, Troy Lee and Ansis Rosmanis for valuable discussions.

This work has been supported by the European Social Fund within the project ``Support for Doctoral Studies at University of Latvia''. 

\bibliographystyle{abbrv}
\bibliography{span_distinct}

\end{document}